\documentclass[aps,reprint,nofootinbib, amsmath,amssymb,showpacs,showkeys,superscriptaddress]{revtex4-1}

\usepackage{graphicx}  
\usepackage{natbib}  
\usepackage{amsmath}  
\usepackage[colorlinks=true,linkcolor=blue,citecolor=blue]{hyperref}  
\usepackage[top=1.5in, bottom=1.5in, left=1in, right=1in]{geometry}
\usepackage{caption}
\usepackage{amsfonts}
\usepackage{xcolor}
\usepackage{booktabs}
\usepackage{multirow}
\usepackage{bbm}
\usepackage{soul,url}
\usepackage{afterpage}

\begin{document}

\title{Generative adversarial network for stellar core-collapse gravitational waves}

\author{Tarin Eccleston} \affiliation{Department of Statistics,
  University of Auckland, Auckland 1010, New Zealand}

\author{Matthew C. Edwards} \affiliation{Department of Statistics,
  University of Auckland, Auckland 1010, New Zealand}

\begin{abstract}

We present a rapid stellar core-collapse waveform emulator built using a deep convolutional generative adversarial network (DCGAN).  The DCGAN was trained on the Richers \textit{et al.~}\cite{richers:2017} waveform catalogue to learn the structure of rotating stellar core-collapse gravitational-wave signals and generate realistic waveforms.   We show that the DCGAN learns the distribution of the training data reasonably well, and that the waveform emulator produces signals that appear to have the key features of core-collapse, bounce, early post-bounce, and ringdown oscillations of the early proto-neutron star.  The pre-trained DCGAN can therefore be used as a phenomenological model for rotating stellar core-collapse gravitational-waves.  


\end{abstract}


\maketitle

\section{Introduction}\label{sec:intro}

\begin{figure}[!h]
    \centering
    \includegraphics[width = 1\linewidth]{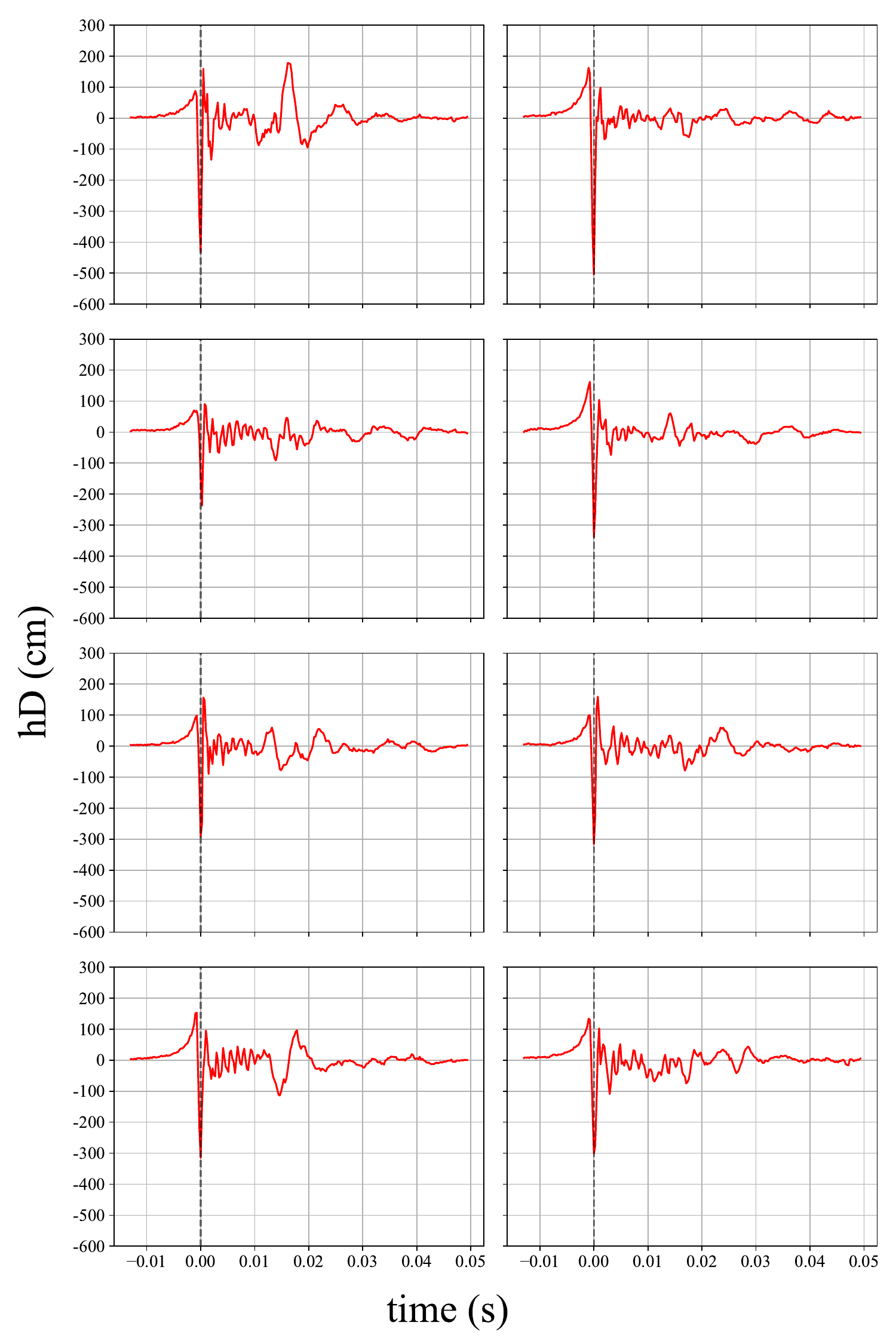}
    \caption{Emulated rotating stellar core-collapse gravitational-wave signals, trained on the Richers \textit{et al.}~\cite{richers:2017} waveform catalogue.}
    \label{fig:fake}
\end{figure}

Figure~\ref{fig:fake} displays a set of emulated rotating stellar core-collapse gravitational-wave signals.  Each signal took an average of $4.633 \times 10^{-3}$ seconds (with standard deviation $5.306 \times 10^{-5}$) to generate using an Apple M2 chip with the Metal Performance Shader (MPS) framework.  The signals were generated using generative deep learning, specifically a deep convolutional generative adversarial network (DCGAN) \cite{goodfellow:2014, radford:2016}.  The DCGAN was trained using the Richers \textit{et al.}~\cite{richers:2017} rotating stellar core-collapse waveform catalogue, and took $587.1$ seconds to train on the same processor.  The pre-trained DCGAN can be thought of as a phenomenological model for rotating core-collapse gravitational waves, mimicking the key features of a rotating stellar core-collapse gravitational-wave signal, exhibiting the collapse, bounce, and early post-bounce and ringdown oscillations of the early proto-neutron star  \cite{dimmelmeier:2008, abdikamalov:2014, richers:2017}.  

Gravitational-waves from Galactic core-collapse supernovae should be observable using current detectors \cite{gossan:2016}.  However, after three full observing runs (O1--O3) and one partial observing run (O4) by the network of Earth-based GW detectors, Advanced LIGO \cite{lsc:2015}, Advanced Virgo \cite{virgo:2015}, and KAGRA \cite{kagra:2020}, gravitational-waves from stellar core-collapse have not yet been observed \cite{ligo_ccsn:2020}.  Gravitational-waves from stellar core-collapse carry information about the core-collapse dynamics, explosion mechanism, evolution of the proto-neutron star, rotation rate, and nuclear equation-of-state \cite{gossan:2016}, providing a direct probe into the core of a collapsing star.

Stellar core-collapse gravitational-wave signals are difficult to model, interfacing gravitational, nuclear, particle, statistical, and numerical physics \cite{vartanyan:2023b}, and taking an order of months to generate using the most cutting-edge waveform modelling \cite{schnetter:2007, ott:2009}.  Therefore, unlike compact binary coalescence (CBC) signals, template-bank search algorithms for detection, such as matched-filtering, have not been possible for stellar core-collapse, and alternative methods have had to be developed for burst-type gravitational-wave events \cite{klimenko:2016, cornish:2015}, which includes stellar core-collapse.

The computational challenges of waveform modelling has also limited the development of parameter estimation routines for stellar core-collapse as inference generally requires evaluating a likelihood that is a function of the signal. Meyer \textit{et al.} \cite{meyer:2022} provide a comprehensive overview of parameter estimation in gravitational-wave astronomy.  

Most of the early parameter estimation literature used principal component regression on waveform catalogues to reconstruct core-collapse signals embedded in noise \cite{heng:2009, roever:2009}.  Bayesian inference has been used to infer the explosion mechanism \cite{logue:2012}, ratio of rotational to potential energy \cite{edwards:2014}, and time evolution of a particular combination of the mass and radius of the protoneutron star \cite{bizouard:2021, bizouard:2023}.  Deep learning has been used to develop a detection pipeline \cite{astone:2018, lopez:2021, iess:2023}, infer the explosion mechanism \cite{chan:2020, iess:2020}, and nuclear equation-of-state \cite{edwards:2021}.

Generative models in deep learning \cite{goodfellow:2016}, such as generative adversarial networks (GANs) \cite{goodfellow:2014}, variational autoencoders (VAEs) \cite{kingma:2022}, and generative pre-trained transformers (GPTs) \cite{vaswani:2017} are drastically changing the world.  We have seen a meteoric rise in the use of text generation platforms such as ChatGPT \cite{radford:2018}, and image generation with latent diffusion models with tools such as DALL-E \cite{shi:2020, rombach:2022}. Generative methods have also propagated into the sciences, with great success in biology, neuroscience, and astrophysics \cite{jorgensen:2018, ingraham:2019, anand:2018, ramezanian:2022, schawinski:2017}. Generative deep learning is now being used in GW astronomy \cite{benedetto:2023}; for performing rapid Bayesian parameter estimation of the compact binary coalescence (CBC) of black holes \cite{green:2020, gabbard:2022}, performing rapid CBC waveform generation necessary for low-latency detection \cite{liao:2021, lee:2021}, and generating burst GWs \cite{mcginn:2021} and glitches \cite{lopez:2022, powell:2023}, useful for testing detection pipelines.  Further, artificial neural networks have been used to create a fast and fully-relativistic waveform generator for extreme-mass ratio inspiral (EMRI) GW signals \cite{chua:2019, chua:2021, katz:2021} and CBC waveforms \cite{grimbergen:2024}.  

A phenomenological model is one that has parameters that are not directly derived from first physical principles.  They are common in gravitational-wave data analysis.  For example, \cite{astone:2018} created a phenomenological model using a harmonic oscillator to simulate core-collapse supernova gravitational waves to match the most common features seen in numerical models of stellar core-collapse.  They used this to derive a template bank.  

In this paper, we also derive a template bank, but based on learning the underlying distribution of the Richers~\textit{et al.} \cite{richers:2017} waveform catalogue.  To our knowledge, this is the first attempt at using generative deep learning to create a rapid waveform emulator and phenomenological model for stellar core-collapse gravitational-waves.  We see this being useful for data augmentation, which can aid in the development of new parameter estimation routines for stellar core-collapse.  We also see this being useful for testing existing detection pipelines \cite{cornish:2015, klimenko:2016}, and the beginnings of matched-filter type analyses for stellar core-collapse gravitational-waves.  

This article is structured as follows.  In Section~\ref{sec:gans} we provide a primer on GANs. We then discuss the training set from the Richers~\textit{et al.}~\cite{richers:2017} waveform catalogue and data pre-processing in Section~\ref{sec:data}. This is followed by a description of the specific network architecture we implemented in Section~\ref{sec:specs}.  We present our findings in Section~\ref{sec:results}, followed by a brief discussion of the possible uses of the DCGAN and future initiatives in Section~\ref{sec:end}.

\section{Generative adversarial networks (GANs)}\label{sec:gans}

A generative adversarial network (GAN) is an unsupervised machine learning algorithm containing two competing neural networks; the \textit{generator} network and the \textit{discriminator} network.  The discriminator learns to discern between real samples drawn from the training set and fake samples drawn from the generator, while the generator starts with a random data distribution and learns a latent representation of the training data.  Informally, the generator tries to trick the discriminator with fake data, the discriminator tries to detect this fake data, and this zero-sum competition allows both networks to improve, leading to more realistic generated data.  GANs are therefore useful for data augmentation, i.e., creating new data based on existing data to better train machine learning and statistical algorithms.  

\begin{figure}
    \centering
    \includegraphics[width = 1\linewidth]{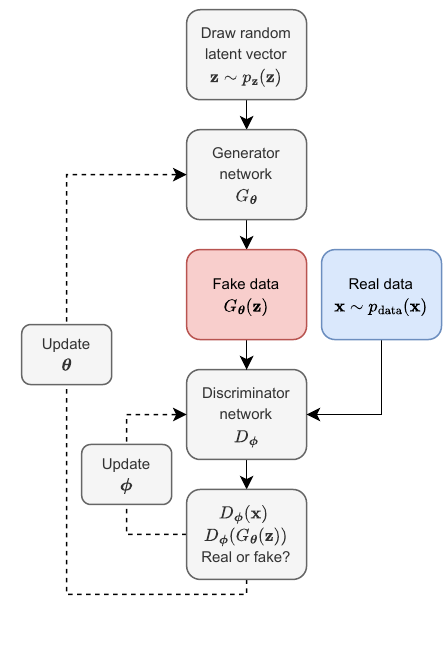}
    \caption{GAN diagram. The discriminator network learns whether input data is real or fake, and the generator network learns to generate realistic fake data.}
    \label{fig:gan_diagram}
\end{figure}

Figure~\ref{fig:gan_diagram} shows the general process for training a GAN.  Let $\mathbf{x}$ be samples from real data in the training set and $\mathbf{z}$ be random samples from a latent probability space (usually multivariate Gaussian).  Let the generator and discriminator networks have the following network parameters 
$(\theta, \phi)$ respectively.  The generator network $G_{\theta}(\mathbf{z})$ generates fake data by mapping random samples $\mathbf{z}$ from the latent probability space to the space of the training set such that $\mathbf{x}$ and $G_{\theta}(\mathbf{z})$ have the same dimensions.  The discriminator network $D_{\phi}(.)$ is a binary classifier that outputs the probability that its input is real, i.e., it predicts whether the input is real or fake.  If $y$ is the true label (where $y = 1$ is the label given to real data and $y = 0$ is the label given to fake data), and $\hat{y}$ is the network prediction, then the discriminator tries to maximise the binary cross-entropy loss function:
\begin{equation*}
L(y, \hat{y}) = y \log(\hat{y}) + (1 - y) \log(1 - \hat{y}),
\end{equation*}
where $\hat{y} = D_{\phi}(\mathbf{x})$ is the network prediction if the data is real and $\hat{y} = D_{\phi}(G_{\theta}(\mathbf{z}))$ is the network prediction if the data is fake.  This leads to the following GAN objective function:
\begin{widetext}
\begin{equation}\label{eq:objective}
    \min_{\theta} \max_{\phi} V(G_{\theta}, D_{\phi}) =    
    \mathbb{E}_{\mathbf{x} \sim p_{\text{data}}(\mathbf{x})}[\log D_{\phi}(\mathbf{x})] +  \mathbb{E}_{\mathbf{z} \sim p_{\mathbf{z}}(\mathbf{z})}[\log(1 - D_{\phi}(G_{\theta}(\mathbf{z}))].
\end{equation}
\end{widetext}

In game theory, this is an example of a two-player non-cooperative minimax game, where the generator minimizes $V(G_{\theta}, D_{\phi})$ with respect to its network parameters $\theta$, and the discriminator maximizes $V(G_{\theta}, D_{\phi})$ with respect to its network parameters $\phi$.  The discriminator wants to maximise $D_{\phi}(\mathbf{x})$ to be close to 1 to give high probability that the real data is real, and to minimise $D_{\phi}(G_{\theta}(\mathbf{z}))$ to be close to 0 to give low probability that the fake data is real.  Since the generator wants to fool the discriminator, the generator wants to maximise $D_{\phi}(G_{\theta}(\mathbf{z}))$ to be close to 1 so that the discriminator gives high probability that the fake data is real.

During training, the network parameters $(\theta, \phi)$ are updated in batches of $m$ real samples from the training set and $m$ fake samples from the generator using backpropagation.  Along with the true labels, the samples are passed into the discriminator to calculate the ``discriminator loss''.  The gradient of the objective function with respect to the discriminator's parameters $\phi$,
\begin{equation*}
  \nabla_{\phi}\frac{1}{m}\sum_{i=1}^m \left( \log(D_{\phi}(\mathbf{x}^{(i)})) + \log(1 - D_{\phi}(G_{\theta}(\mathbf{z}^{(i)}))) \right),
\end{equation*}
is then calculated to guide the neural network in which direction to update $\phi$ to maximise Equation~\ref{eq:objective} (gradient ascent).  The $m$ fake samples from the generator are then used to calculate the ``generator loss''.  The gradient of the objective function with respect to the generator's parameters $\theta$,
\begin{equation*}
  \nabla_{\theta}\frac{1}{m}\sum_{i=1}^m \left(\log(1 - D_{\phi}(G_{\theta}(\mathbf{z}^{(i)}))) \right),
\end{equation*}
is then calculated to determine the direction in which to change $\theta$ to minimise Equation~\ref{eq:objective} (gradient descent).

\subsection{Multilayer perceptrons (MLPs)}

The original GAN implementation by Goodfellow~\textit{et al.}~\cite{goodfellow:2014} used multilayer perceptrons (MLPs) for the generator and discriminator networks.  An MLP is the standard deep feed-forward artificial neural network with layers of fully-connected neurons.  This means that each neuron, $j = 1, 2, \ldots, m$, in a given layer may have many inputs, $(x_1, x_2, \ldots, x_n)$, and one output, $y_j$. Each input has a weight  $(w_{1,j}, w_{2,j}, \ldots, w_{n,j})$ and a neuron may have a bias $w_{0,j}$ associated with another input $x_0 = 1$. The weights and bias are the learnable parameters of each neuron. The neuron is activated by computing the linear combination of the inputs and weights/biases and then applying a non-linear activation function $f(.)$ to compute its output $y_j$. That is, for each neuron, $j = 1, 2, \ldots, m$, in a given layer,
\begin{eqnarray*}
    a_j &=& \sum_{i=0}^n w_{i,j} x_i \\
    y_j &=& f(a_j).
\end{eqnarray*}
The neurons in the input layer $(x_1, x_2, \ldots, x_n)$ and the neurons in the output layer $(y_1, y_2, \ldots, y_m)$ are then said to be fully-connected.  If there is more than one hidden layer, then the neural network is an MLP, and the more layers there are, the deeper the MLP.

\subsection{Convolutional neural networks (CNNs)}

Rather than using MLPs for the generator and discriminator networks, convolutional neural networks (CNNs) can be used instead to implement a deep convolutional GAN (DCGAN) \cite{radford:2016}.  Rather than using fully-connected layers, DCGANs use convolutional layers.  For our work, one-dimensional convolutional layers (\texttt{Conv1D}) are used as they can encode abstract temporal features in sequence data, making them useful for learning patterns in time series.  

A \texttt{Conv1D} layer consists of a set of learnable filters that slide along its input layer, computing the convolution which expresses the amount of overlap between the input and the filter.  For input sequence $x$ and filter $k$, a one-dimensional convolution is defined as:
\begin{equation*}
    (x \ast k)(t) = \sum_{s = -\infty}^{\infty} x(s) k(t - s),
\end{equation*}
for times $s$ and $t$.  A \texttt{Conv1D} layer applies this convolution over all filters in the layer.   A \texttt{ConvTranspose1D} layer applies the reverse (upsampling) transformation.  

The number and size of the filters are user-defined, where the number of filters tells the neural network how many features to learn, and the size of the filter determines the size of the features in the input that the layer can learn.  The amount that the filters are moved by is called the stride, and larger stride values provide larger amounts of compression.  To preserve the length of the input after convolution, zero padding can be applied.  The output of the layer is called a feature map.  The filters are the trainable parameters under this neural network architecture. 

\subsection{Useful activation functions}

A commonly used activation function in deep learning is LeakyReLU, which is defined as:
\begin{equation*}
    f(x)= 
\begin{cases}
    x,& \text{if } x\geq 0\\
    \alpha x,              & \text{otherwise},
\end{cases}
\end{equation*}
for slope $\alpha \in (0, 1)$.  If $\alpha = 0$, this reduces to the original rectified linear unit (ReLU).  ReLU is a commonly used activation function to introduce nonlinearity into the neural network and to help mitigate the vanishing gradient problem during training.  LeakyReLU allows for non-zero values when $x < 0$ and mitigates the ``dying neuron problem'' where some neurons in a neural network never get activated.  

The sigmoid activation function is often used for binary classification problems as it maps the real number line to the unit interval.  This is defined as:
\begin{equation*}
    f(x) = \frac{1}{1 + \exp(-x)}, \quad x \in \mathbb{R}.
\end{equation*}

\subsection{Training DCGANs}

GANs and DCGANs are difficult to train and can suffer from the vanishing gradient problem and mode collapse.  

The vanishing gradient problem is when the gradients of the loss function become close to zero and are backpropagated through the neural network \cite{basodi:2020}.  This leads to unstable training and slow convergence, where network parameters can no longer update.  For DCGANs, this can occur early in training when the generator produces poor fake data that the discriminator can easily discern it from real data.  There are two common fixes to this problem: modifying the objective function, i.e., Wasserstein loss \cite{arjovsky:2017} or non-saturating generator loss where the generator maximises $\log(D_{\phi}(G_{\theta}(\mathbf{z}))$ rather than minimising $\log(1 - D_{\phi}(G_{\theta}(\mathbf{z}))$ \cite{goodfellow:2014}, and limiting the depth and complexity of the neural network.

Mode collapse occurs when the generator network gets stuck at a local mode and only produces a small variety of samples, missing many of the modes of the training data distribution  \cite{metz:2016}.  This can happen when the generator learns faster than the discriminator and is rewarded too often for fooling the discriminator.  A Wasserstein loss function can mitigate mode collapse by stabilising training, as well as other regularisation techniques, such as one-sided label smoothing \cite{salimans:2016, szegedy:2015}, where labels can include a random element.  Visual assessment of generated data can be used to inform if mode collapse occurs.  

Other best practices for improving the training of DCGANs include batch normalisation, and spatial dropout.  Batch normalisation is the process of normalising layer inputs in mini-batches to address the problem of internal covariate shift \cite{ioeff:2015}.  This helps increase the learning rate and reduce the number of epochs required in training a neural network.  Spatial dropout is a regularisation technique to prevent over-fitting and stabilise learning in a neural network, where connections between nodes are severed with probability, $p$.  This reduces reliance on the network using particular neurons to learn patterns.  

To assess convergence of a DCGAN, one must monitor the discriminator and generator gradients to ensure they do not explode or vanish, and loss functions to ensure the discriminator loss converges to 0.5, where the discriminator can no longer discern between real and fake samples.  One can also use distance metrics such as the Jensen-Shannon divergence to monitor how closely the generated data distribution matches the training data distribution, as the minimum of the objective function in Equation~\ref{eq:objective} is achieved if and only if the probability distribution of the generator matches the probability distribution of the training data \cite{goodfellow:2014}.

\section{Training set} \label{sec:data}

The data set we used to train the DCGAN was the Richers~\textit{et al.}~\cite{richers:2017} waveform catalogue, publicly available on Zenodo \cite{richers-data:2016}.  This catalogue contains 1824 axisymmetric general-relativistic hydrodynamic simulated rotating stellar core-collapse and bounce gravitational-wave signals.  The signals assume the magnetorotational explosion mechanism, where the star explodes by way of bipolar jet-like outflows driven by rapid rotation and strong magnetic fields.  All of the signals in the catalogue assume an event distance of 10~kpc, i.e., galactic stellar core-collapse.  

The Richers~\textit{et al.}~\cite{richers:2017} waveform catalogue was originally created to probe the effects of the nuclear equation of state and electron capture rates, with varying rotation profiles, on the gravitational-wave signature.  Due to its relatively large size, this is a rich data set for testing machine learning algorithms, such as CNNs to classify the equation of state in \cite{edwards:2021, mitra:2024}.  We chose this catalogue as our training set for this reason.  

Some of the signals in this catalogue do not collapse/bounce due to large centrifugal support from the rapid rotation.  These signals have a different waveform morphology to those that do collapse/bounce and do not emit any significant gravitational-waves, so we excluded them from our analysis.  Once we removed these signals, we were left with 1684 signals for our training set.  

To prepare the training data for the DCGAN, we pre-processed it.  The original sampling rate of the signals was $2^{16}~\text{Hz}$.  We downsampled this to $2^{12}~\text{Hz}$ to reduce the data volume and because the network of second generation terrestrial gravitational-wave observatories will have low sensitivity at high frequencies.  We downsampled by multiplying the data in the time-domain by a Tukey window (with tapering parameter $\alpha = 0.1$), applying an order 10 Butterworth filter (with attenuation 0.25) and removing data using linear interpolation digital resampling \cite{smith:1984}.  This approach mitigates spectral leakage and aliasing.  

We then restricted the domain of the signal to be $(t_b - 12.80, t_b + 49.55)~\text{ms}$ where $t_b$ is the time of bounce.  We did this such that each signal had length $n = 256$.  This allowed us to capture the key components of the signals (collapse, bounce, early pre-bounce, ringdown oscillations of the PNS).  The gravitational-wave signal has little energy before 2~ms before bounce \cite{dimmelmeier:2008, mitra:2024}, so the lower truncation is reasonable. However, prompt-convection occurs after around 6~ms and because this is stochastic \cite{dimmelmeier:2008}, one gravitational-wave simulation per initial condition setting cannot capture all possible convection dynamics \cite{mitra:2024}, and the quality of the signals is therefore less reliable after 6~ms.  Regardless of this limitation we are interested in seeing the evolution of generated signals during prompt-convection and ringdown.  We therefore do not truncate the training data any further.  

Further, the signals are scaled before inputting them into the DCGAN to mitigate numerical issues.  Each signal in the training set is divided by the maximum absolute strain times distance (where distance is set to 10~kpc in cm) over the entire catalogue.

\section{Network architecture}\label{sec:specs}

\subsection{Generator network}

The neural network architecture for our generator is given in Table~\ref{tab:generator_architecture}.  The generator network starts by randomly sampling the latent vector $\mathbf{z}$ from a 256-dimensional Gaussian distribution with zero mean vector and identity covariance matrix.  We then apply seven \texttt{ConvTranspose1D} layers with batch normalisation such that the number of filters progressively decreases and the filter length increases until the final output is a length $256$ time series; the same dimension as the training data. This neural network has $\sim 13,280,000$ learnable parameters.

\begin{table}[!h]
\centering
\begin{tabular}{cccc}
\hline
Layer & Type & Output Shape & Activation \\
\hline
0 & Input & (256, 1) &  \\
1 & ConvTranspose1D & (2048, 4) & \\
2 & BatchNorm1D & (2048, 4) & LeakyReLU \\
3 & ConvTranspose1D & (1024, 8) & \\
4 & BatchNorm1D & (1024, 8) & LeakyReLU \\
5 & ConvTranspose1D & (512, 16) & \\
6 & BatchNorm1D & (512, 16) & LeakyReLU \\
7 & ConvTranspose1D & (256, 32) & \\
8 & BatchNorm1D & (256, 32) & LeakyReLU \\
9 & ConvTranspose1D & (128, 64) & \\
10 & BatchNorm1D & (128, 64) & LeakyReLU \\
11 & ConvTranspose1D & (64, 128) & \\
12 & BatchNorm1D & (64, 128) & LeakyReLU \\
13 & ConvTranspose1D & (1, 256) & Identity \\
\hline
\end{tabular}
\caption{Generator network architecture with 7 up-sampling convolutional layers.  The LeakyReLU parameter is set to 0.01.}
\label{tab:generator_architecture}
\end{table}

Along with using the LeakyReLU activation function (with $\alpha = 0.01$), batch normalisation is also applied to allow for healthy gradient flow when training \cite{dcgans_pytorch_tutorial}.  No dropout is used for the generator network as randomness is introduced via the latent probability space.

\subsection{Discriminator network}

The neural network architecture for the discriminator can be seen in Table~\ref{tab:discriminator_architecture}.  The input is a length $256$ vector; either a real or generated signal. We apply four \texttt{Conv1D} layers with batch normalisation and dropout.  Each convolutional layer outputs a feature map that doubles the number of filters while halving the filter size.  The activation function used on these layers is LeakyReLU, with $\alpha = 0.2$.  We then flatten the network and apply one fully-connected layer.  On the final layer, we apply a sigmoid activation function to return a value in $[0, 1]$ representing the probability that the input into the discriminator is a real signal.  The number of learnable parameters in the discriminator network is $\sim 698,000$

\begin{table}[!h]
\centering
\begin{tabular}{cccc}
\hline
Layer & Type & Output Shape & Activation \\
\hline
0 & Input & (1, 256) &  \\
1 & Conv1D & (64, 128) & LeakyReLU \\
2 & Dropout1D & (64, 128) &  \\
3 & Conv1D & (128, 64) & \\
4 & BatchNorm1D & (128, 64) & LeakyReLU \\
5 & Dropout1D & (128, 64) &  \\
6 & Conv1D & (256, 32) & \\
7 & BatchNorm1D & (256, 32) & LeakyReLU \\
8 & Dropout1D & (256, 32) &  \\
9 & Conv1D & (512, 16) & \\
10 & BatchNorm1D & (512, 16) & LeakyReLU \\
11 & Dropout1D & (512, 16) &  \\
\hline
Layer & Type & \# Output Neurons & Activation \\
\hline
12 & Flatten & 8192 & \\
13 & Fully-Connected & 1 & Sigmoid \\
\hline
\end{tabular}
\caption{Discriminator network architecture with 4 down-sampling convolutional layers and 1 fully-connected layer. The dropout parameter is set to 0.2, and the LeakyReLU parameter is set to $\alpha = 0.2$.}
\label{tab:discriminator_architecture}
\end{table}

\subsection{Other specifications}

For all of the \texttt{Conv1D} and \texttt{ConvTranspose1D} layers, we use a filter size of 4, stride of 2, padding of 1 (apart from the first layer in the generator, which uses no padding).  The stride hyperparameter allows the neural networks to learn how to pool \cite{radford:2016}.

To improve training we used non-saturating loss \cite{goodfellow:2014}, and one-sided label smoothing on the discriminator \cite{salimans:2016}, setting real labels to be $1.0$ and fake labels to be a random uniform value between $0.0$ and $0.25$.

We implemented the DCGAN using \texttt{PyTorch}, training for 128 epochs using an Adam optimiser with learning rate of 0.00002 for both neural networks, and coefficients for computing running averages of the gradient, $\boldsymbol\beta = (0.5, 0.99)$.  We set the batch size to be 32 and monitored generator and discriminator loss and gradients for each batch to ensure the algorithm was training well and not suffering from mode collapse or vanishing gradients.

\section{Results}\label{sec:results}

We saw in Figure~\ref{fig:fake} that the DCGAN can mimic the key features of rotating stellar core-collapse gravitational-waves.  We provide some further analysis in this section.

\subsection{Comparing distributions}

\begin{figure}[!h]
    \centering
    \includegraphics[width = 1\linewidth]{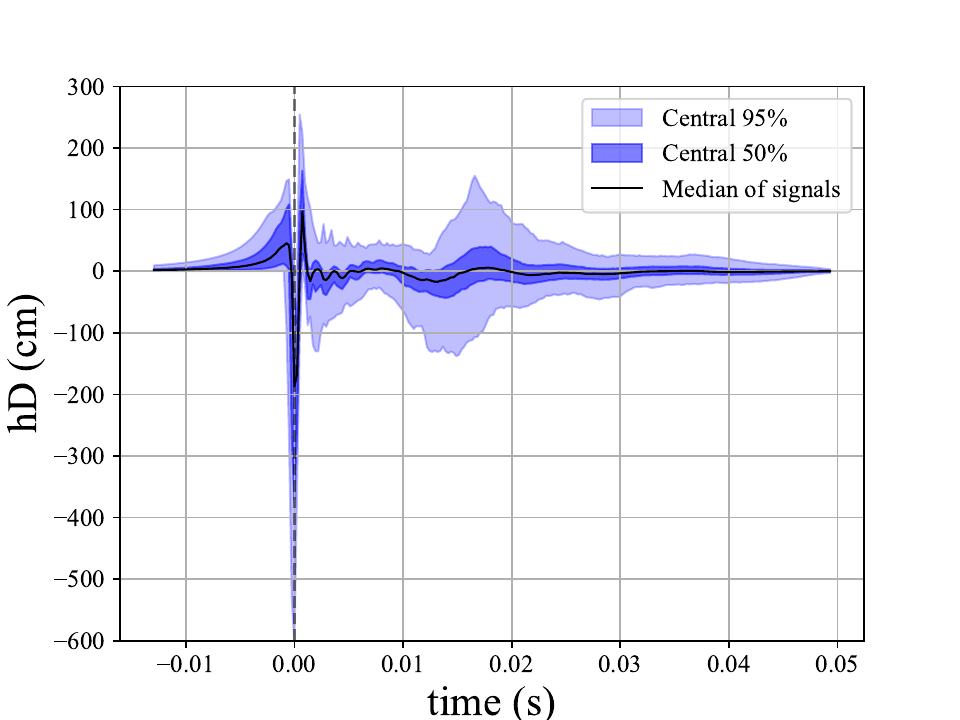}
    \includegraphics[width = 1\linewidth]{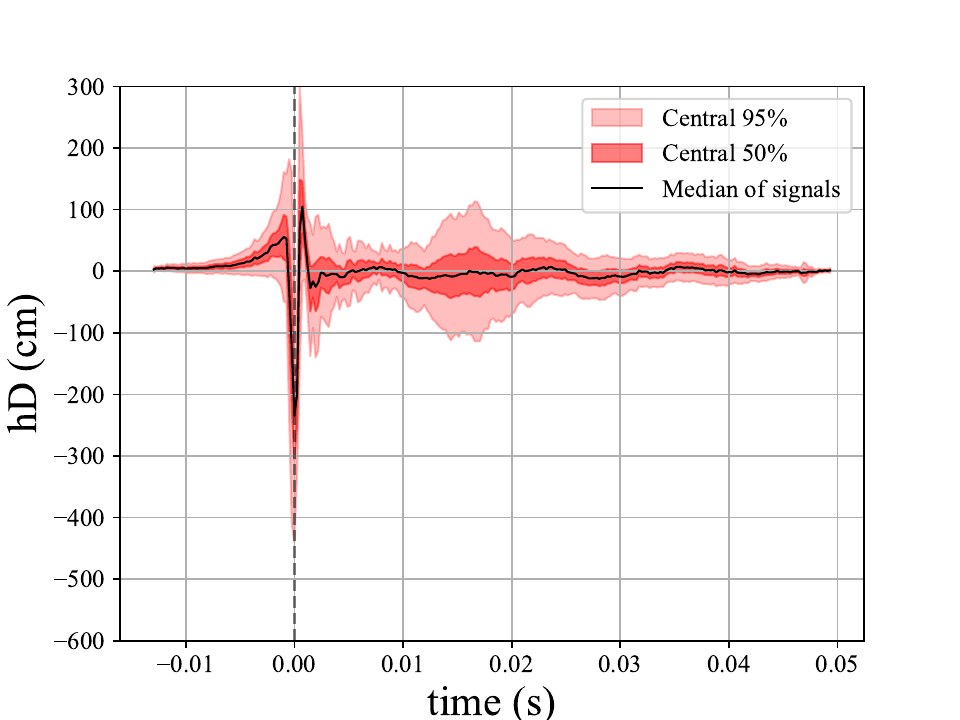}
    \caption{Distribution of real (blue) and emulated (red) rotating stellar core-collapse gravitational-wave signals.  The real signals are the pre-processed Richers \textit{et al.}~\cite{richers:2017} catalogue signals (excluding those that did not collapse). The 10,000 emulated signals come from the DCGAN generator that was trained on the real signals from Richers \textit{et al.}~\cite{richers:2017}.}
    \label{fig:distributions}
\end{figure}

In Figure~\ref{fig:distributions}, we plot the median, central 50\%, and central 95\% of the gravitational-wave signals in the real (blue) and fake (red) data sets.  The emulated signals mimic the key features of the real rotating core-collapse signals, including the collapse, bounce, and early post-bounce stages.  However, there are some notable differences between the two data sets.  First, the real signals are strictly positive before core bounce (which occurs around $t = 0$), whereas the emulated signals have negative values in the tails of the distribution, which is not astrophysically meaningful.  Second, the real signals have a skewed distribution for many of the time points, such as in the early post-bounce stage where the median is not in the center of the central 50\%, whereas the emulated signals appear to have symmetric distributions around the median.  Third, the distribution over time of the signals is a lot smoother for the real signals compared to the emulated signals.  All of these key differences could be due to limited training data, and will investigated in future research.

To measure how dissimilar the probability distributions of the training data and the generated data are at each time index $t = 1, 2, \ldots, 256$, we compute the Jensen-Shannon divergence.  The Jensen-Shannon divergence is a symmetric and smoothed version of the Kullback-Leibler divergence, defined on $[0, 1]$, where a divergence of 0 means that the two probability distributions are identical.  The Jensen-Shannon divergence is defined as:
\begin{equation}
    \text{JSD}(P || Q) = \frac{1}{2} \left(D_{\text{KL}}(P || M) + D_{\text{KL}}(Q || M)\right),
\end{equation}
where $P$ and $Q$ are probability distributions, $M~=~\frac{1}{2}(P~+~Q)$, and $D_{\text{KL}}(P || Q)$ is the Kullback-Leibler divergence for $P$ and $Q$ given by:
\begin{equation}
    D_{KL}(P || Q) = \sum_{\forall{x}} P(x) \log\left(\frac{P(x)}{Q(x)}\right).
\end{equation}

\begin{figure}[!h]
    \centering
    \includegraphics[width = 1\linewidth]{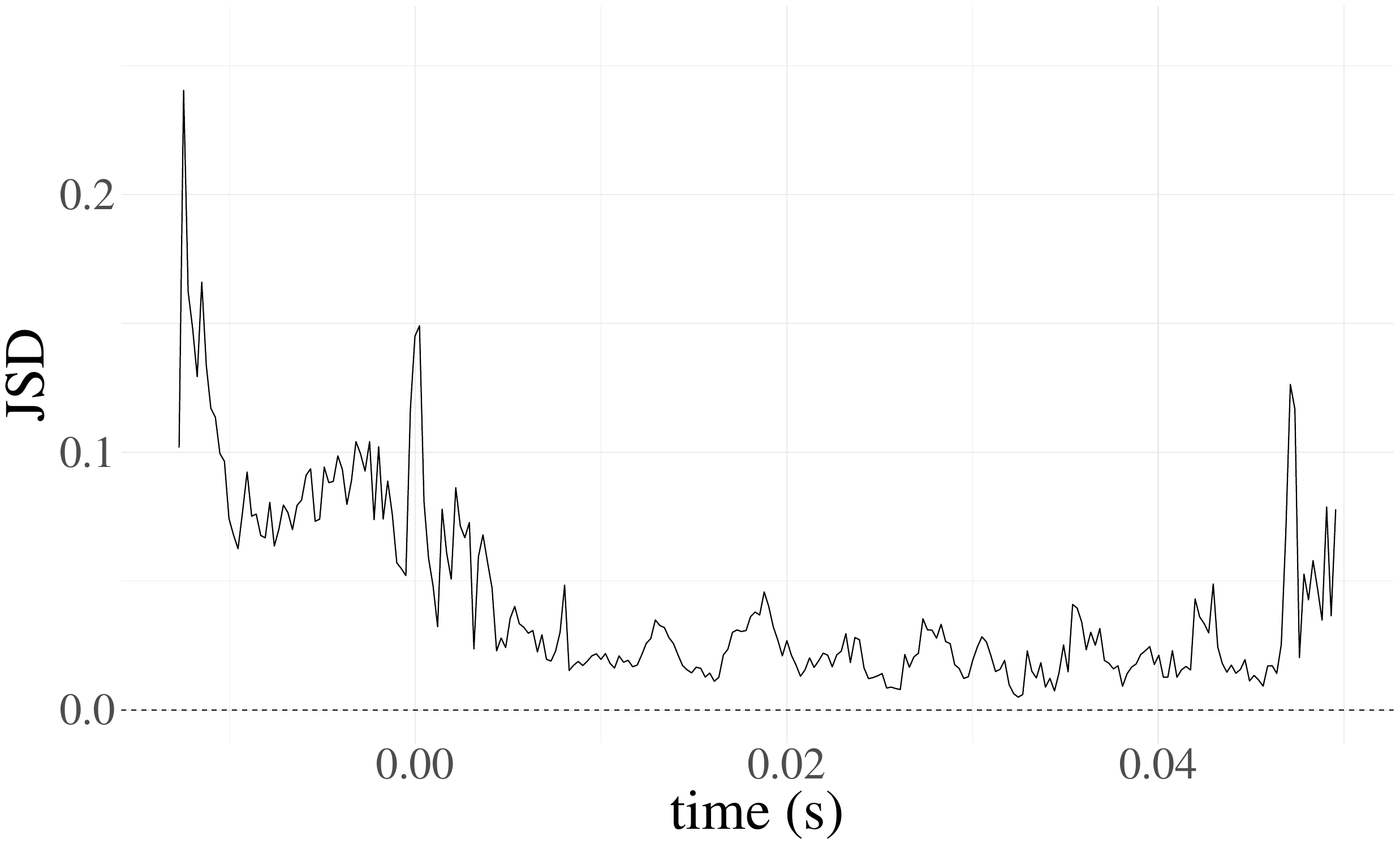}
    \caption{Jensen-Shannon divergence for each time index.}
    \label{fig:JSD}
\end{figure}

In Figure~\ref{fig:JSD}, we show the pointwise Jensen-Shannon divergence, comparing the distribution of real and emulated signals, for each time index.  The divergence was calculated using histograms with 20 bins of equal length over the range of the combined set of real and fake signals.  We can see that the distributions between the real and fake data sets are dissimilar at the start and end of the signals, as well as at core bounce.  We also observe that the distributions are relatively similar between core bounce and the end of the time series.  

It should be noted here that this approach of calculating the point-wise Jensen-Shannon divergence for each time index does not account for time-dependence of the real/generated signals (which is modelled with convolutional layers for the generated signals), and thus provides the worst-case scenario when comparing distributions.  

\begin{figure}[!h]
    \centering
    \includegraphics[width = 1\linewidth]{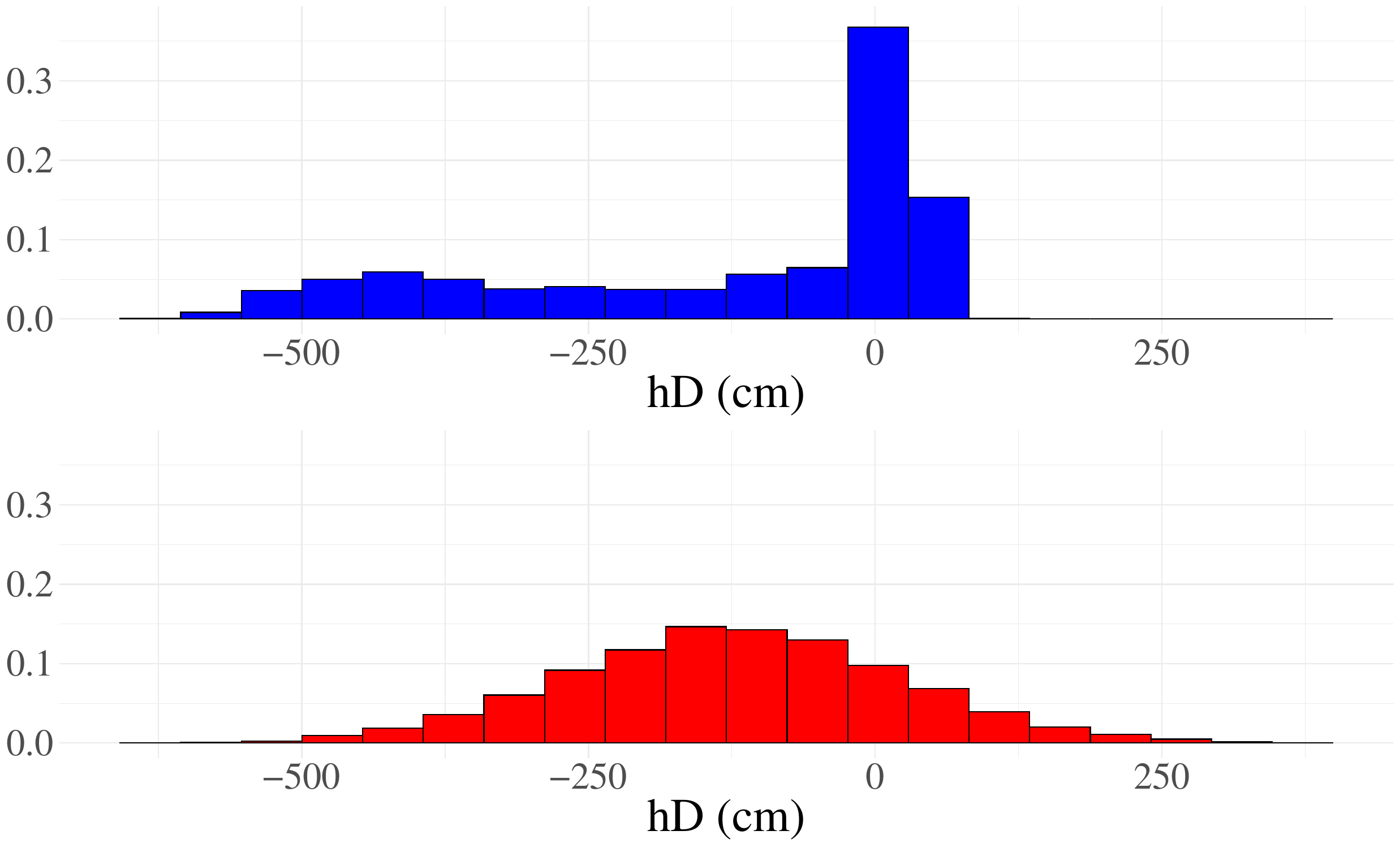}
    \caption{Distribution of GW strain at $t = 0$~s (i.e., the time of core bounce) for real (blue) and fake (red) signals. 
 The Jensen-Shannon divergence is calculated using these histograms, and is equal to 0.149.}
    \label{fig:JSD_Histogram_Bounce}
\end{figure}

In Figure~\ref{fig:JSD_Histogram_Bounce}, we compare the distributions of the GW strain (multiplied by distance) for the real and emulated signals at the time of core bounce (just after $t = 0$~s).  Although the range of the two data sets is similar, the distributions are dissimilar, leading to a relatively large Jensen-Shannon divergence of 0.149.  The distribution of the emulated signals appears to be Gaussian (and this is the case for all time indices), whereas the distribution of the real signals is multimodal and skewed.

\begin{figure}[!h]
    \centering
    \includegraphics[width = 1\linewidth]{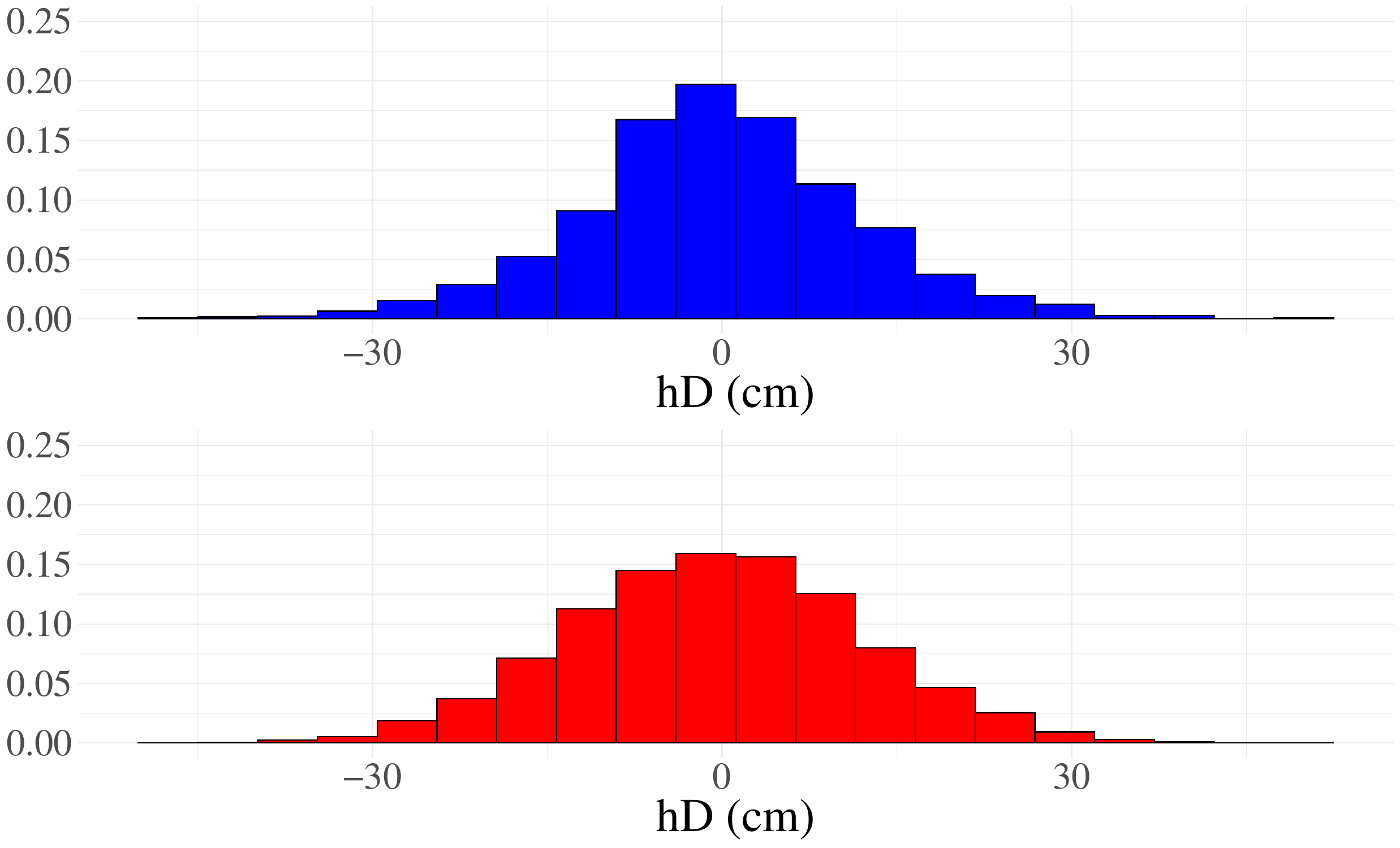}
    \caption{Distribution of GW strain at $t = 0.0325$~s for real (blue) and fake (red) signals. This has the smallest Jensen-Shannon divergence at 0.0049.}
    \label{fig:JSD_Histogram_Min}
\end{figure}

In Figure~\ref{fig:JSD_Histogram_Min}, we see the distributions of GW strain (times distance) at $t = 0.0325$~s.  This corresponds to the time with the lowest Jensen-Shannon divergence, calculated at 0.0049.  We see the distributions are similar, unimodal, and symmetric.  

Although the there are some differences between the distributions of the real and fake signals, it is important to note that the majority of generated signals have learned the time-dependence and general structure of rotating core-collapse gravitational-waves, as seen in Figure~\ref{fig:fake}.  This is due to the convolutional layers learning the temporal structure of the training data.  

\begin{figure}[!h]
    \centering
    \includegraphics[width = 1\linewidth]{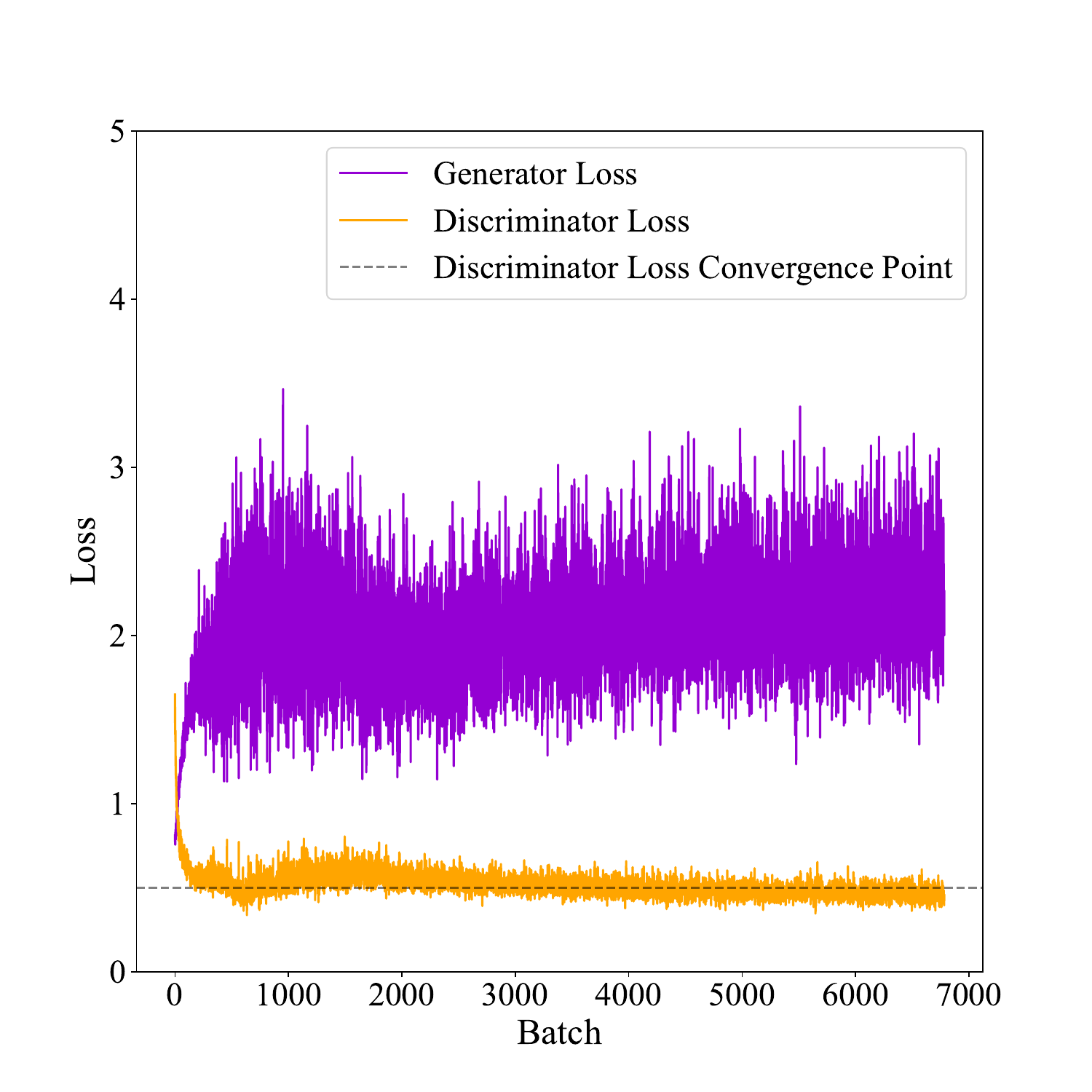}
    \caption{Training loss of discriminator and generator networks, across 1684 signals with a batch size of 32, and 128 training epochs. The discriminator loss has converged to 0.5.}
    \label{fig:training_loss}
\end{figure}

To monitor convergence of the DCGAN, we plot the training loss of the discriminator and generator networks.  Goodfellow~\textit{et al.}~\cite{goodfellow:2014} prove that the optimal discriminator network converges to 0.5.  We see in Figure~\ref{fig:training_loss} that the discriminator converges to this value after 128 epochs.  Note that the number of total batches is equal to $\lceil1684 / 32\rceil \times 128 = 6784$.

\subsection{Evaluating the DCGAN}

Evaluating GANs is an active area of research \cite{borji:2022}. The most common methods for evaluating GANs involve computing either an Inception Score (IS) \cite{salimans:2016} or Fr\'{e}chet Inception Distance (FID) \cite{heusel:2017}.  Both rely on using the existing InceptionNet classifier that was trained on ImageNet.  These are thus very useful for determining a category for images.  As we have time series data with only one category (i.e., our GAN is not a conditional GAN), we approach the problem of evaluating the DCGAN a different way.  We withhold 10\% of pre-processed Richers \textit{et al} \cite{richers:2017} data as a validation set, keeping 90\% for training the DCGAN.  

\begin{figure}[!h]
    \centering
    \includegraphics[width = 1\linewidth]{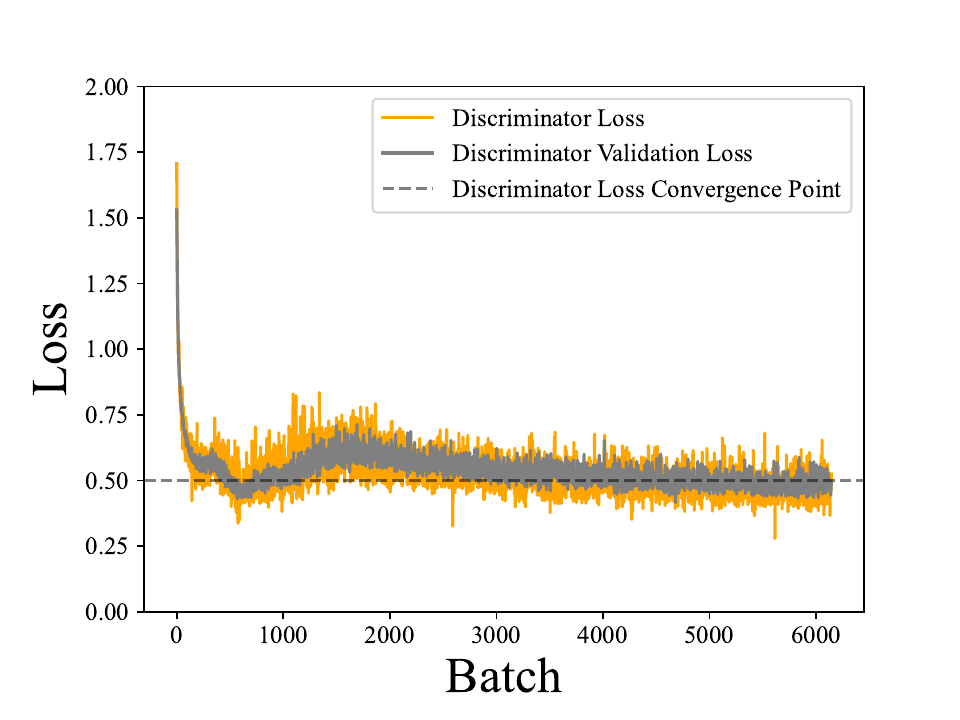}
    \caption{Discriminator loss for training and validation sets.}
    \label{fig:validation_loss}
\end{figure}

To evaluate the performance of the discriminator network, we monitor the discriminator loss for both the training and validation sets at the end of each batch, noting that network parameters are only updated using the training set, and not the validation set.  As seen in Figure~\ref{fig:validation_loss}, both training and validation sets have similar loss trajectories, reaching convergence of 0.5 by the end of the 128 epochs.  This highlights that the discriminator is behaving desirably on core-collapse gravitational-wave signals that it has not been trained on, where it cannot discern between real samples from the validation set and fake samples from the generator. 

\begin{figure}[!h]
    \centering
    \includegraphics[width = 1\linewidth]{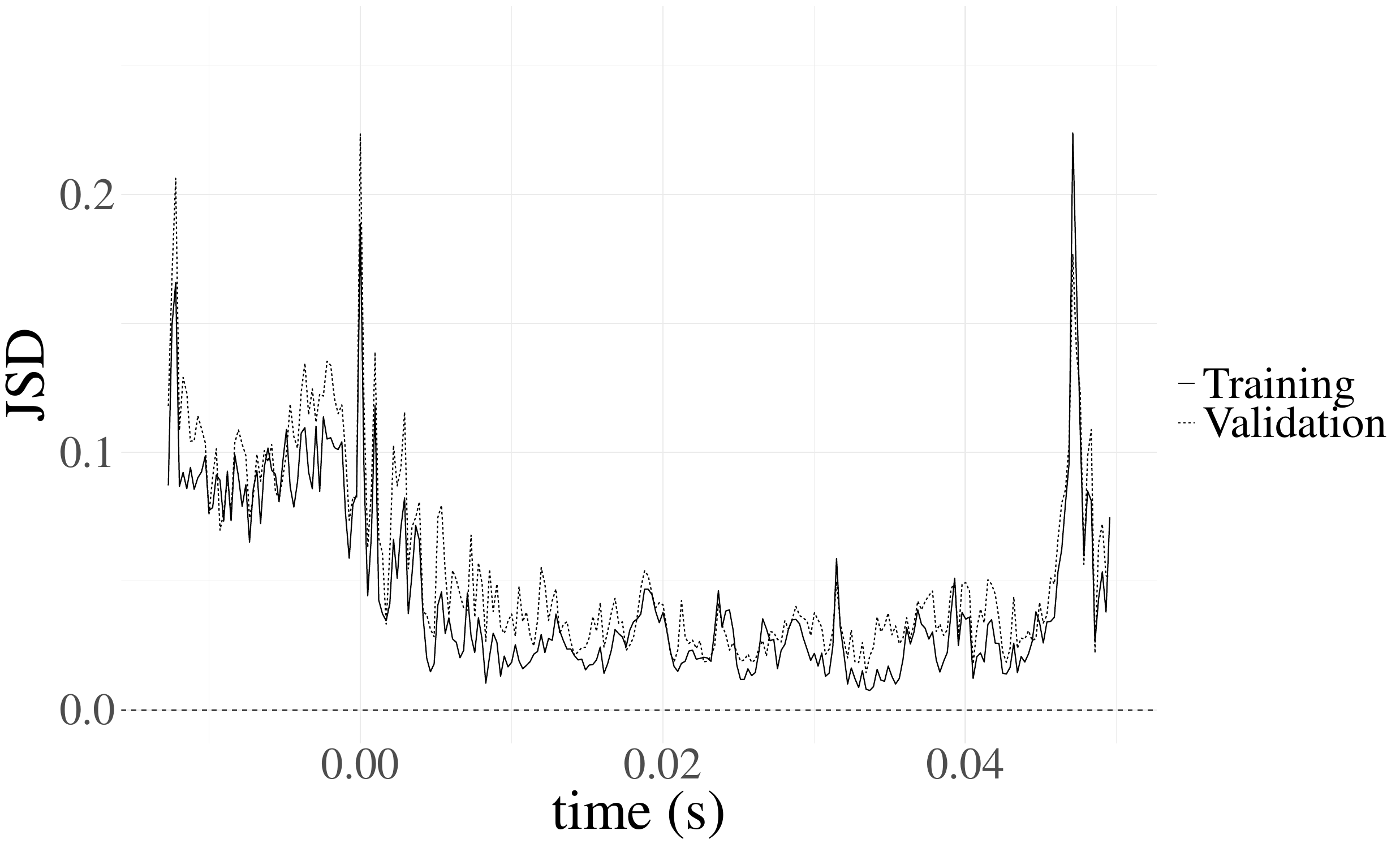}
    \caption{Jensen-Shannon divergence for each time index when comparing the generated signal distribution to the validation set distribution (dashed) and training set distribution (solid).}
    \label{fig:validation_jsd}
\end{figure}

To evaluate the performance of the generator network, we compute the pointwise Jensen-Shannon divergence, comparing the distribution of the validation set and training set to the distribution of fake signals generated by the DCGAN.  We can see in Figure~\ref{fig:validation_jsd} that the pointwise Jensen-Shannon divergence for the validation set behaves similarly to the training set, but with slightly higher values at most time points. This is not unexpected, as the DCGAN is not trained specifically to learn the distribution of the validation set, and the validation set contains fewer signals that are not as diverse as the training set.  However, it is encouraging to see similar behaviours between data sets.  

\section{Starccato and future research}\label{sec:end}

\texttt{Starccato} \footnote{\texttt{https://starccato.github.io/starccato/}} is an open-source \texttt{Python} package.  In its current form, \texttt{Starcatto} allows users to rapidly generate fake but realistic stellar core-collapse gravitational-wave signals from our pre-trained DCGAN.  Currently, the package will generate signals of length $n = 2^8$ at a sampling rate of $2^{12}~\text{Hz}$.  Users who require different specifications or want to use a different training set can easily clone the code, edit it, and retrain the DCGAN.  

As the emulated signals from \texttt{Starcatto} mimic the key features of rotating stellar core-collapse gravitational-wave signals, they can be used as a phenomenological model.  Although not demonstrated in this research, these signals will be useful for data augmentation and the development of new statistical and machine learning approaches in the field of gravitational-wave data analysis; a field that has been hindered by slow waveform generation. \texttt{Starcatto} could also be used to test existing burst pipelines, or generate a template bank for searches for stellar core-collapse gravitational-waves.  This will be shown in a forthcoming article.  

There are many promising directions the \texttt{Starcatto} project can go. One exciting follow-up is exploring conditional DCGANs (cDCGANs) \cite{mirza:2014}, where we train the DCGAN, conditioning it on astrophysical parameters such as the ratio of rotational kinetic energy to gravitational potential energy of the inner core at bounce, pre-collapse differential rotation, nuclear equation-of-state, or explosion mechanism.  McGinn~\textit{et al.} \cite{mcginn:2021} has already shown the usefulness of cDCGANs for generating general burst signals.  In our application, a well-trained cDCGAN will give users the ability to generate rotating core-collapse gravitational-wave signals under certain initial conditions, improving the motivation for a future matched-filter pipeline.  
 
Some of the mismatch between the training set and our generated signals could be explained by the specific network architecture we chose.  In future research, we will explore alternative deep learning architectures to generate realistic core-collapse gravitational-wave signals.  A recurrent neural network (RNN) framework, using RNNs instead of CNNs as the generator and discriminator networks \cite{esteban:2017}, could improve the generator, as RNNs are specifically designed for time series data.  Gated recurrent units (GRUs) \cite{cho:2014, chung:2014} are a promising RNN architecture that uses gating mechanisms to regulate the flow of information, learning long-term time dependencies.  They avoid vanishing gradients, and empirically have faster processing speeds and better performance on smaller training sets than long short term memory (LSTM) networks \cite{yang:2020}.  Transformers \cite{vaswani:2017, wen:2023}, based on multi-head attention, are the current state-of-the-art algorithms when it comes to sequence and time series data.  However, the attention operator in transformers has quadratic cost as a function of sequence length, so limits the amount of context available.  Some researchers consider promising attention-free architectures based on parameterized long convolutions and state-space models \cite{fu:2023a, fu:2023b, poli:2023}. Other researchers consider parameter-efficient fine tuning (PEFT) \cite{xu:2023} that adapts pre-trained transformers to specific tasks, leveraging pre-trained parameters while also including a small number of new domain-specific parameters.  This is relatively computationally cheap and particularly useful for situations where there is limited training data.  Another promising alternative to GANs are variational autoencoders (VAEs) \cite{kingma:2022}, which provide a more interpretable latent space representation by assuming a specific prior distribution. They also allow us to perform variational inference if required, and are less prone to mode collapse and unstable training.  

Suffice to say, there are many exciting avenues to explore when it comes to generating realistic core-collapse gravitational-wave signals with deep learning algorithms, and we believe the research presented in this article is a valuable first step.

\section*{Acknowledgements}

We thank Avi Vajpeyi for his software wizardry.  We also thank Nelson Christensen, Marie-Anne Bizouard, and Ernazar Abdikamalov for their astrophysics expertise.  M. C. E. gratefully acknowledges support by the Marsden Fund Council grant MFP-UOA2131 from New Zealand Government funding, administered by the Royal Society Te Ap\={a}rangi.

\bibliographystyle{unsrt}
\bibliography{refs}  

\begin{thebibliography}{10}

\bibitem{richers:2017}
Sherwood Richers, Christian~D. Ott, Ernazar Abdikamalov, Evan O'Connor, and Chris Sullivan.
\newblock Equation of state effects on gravitational waves from rotating core collapse.
\newblock {\em Phys. Rev. D}, 95:063019, Mar 2017.

\bibitem{goodfellow:2014}
Ian~J. Goodfellow, Jean Pouget-Abadie, Mehdi Mirza, Bing Xu, David Warde-Farley, Sherjil Ozair, Aaron Courville, and Yoshua Bengio.
\newblock Generative adversarial networks.
\newblock {\em arXiv preprint arXiv:1406.2661}, 2014.

\bibitem{radford:2016}
Alec Radford, Luke Metz, and Soumith Chintala.
\newblock Unsupervised representation learning with deep convolutional generative adversarial networks.
\newblock {\em arXiv preprint arXiv:1511.06434}, 2015.

\bibitem{dimmelmeier:2008}
Harald Dimmelmeier, Christian~D. Ott, Andreas Marek, and H.-Thomas Janka.
\newblock Gravitational wave burst signal from core collapse of rotating stars.
\newblock {\em Phys. Rev. D}, 78:064056, Sep 2008.

\bibitem{abdikamalov:2014}
Ernazar Abdikamalov, Sarah Gossan, Alexandra~M. DeMaio, and Christian~D. Ott.
\newblock Measuring the angular momentum distribution in core-collapse supernova progenitors with gravitational waves.
\newblock {\em Phys. Rev. D}, 90:044001, Aug 2014.

\bibitem{gossan:2016}
S.~E. Gossan, P.~Sutton, A.~Stuver, M.~Zanolin, K.~Gill, and C.~D. Ott.
\newblock Observing gravitational waves from core-collapse supernovae in the advanced detector era.
\newblock {\em Phys. Rev. D}, 93:042002, Feb 2016.

\bibitem{lsc:2015}
The LIGO~Scientific Collaboration, J~Aasi, B~P Abbott, et~al.
\newblock Advanced {LIGO}.
\newblock {\em Classical and Quantum Gravity}, 32(7):074001, mar 2015.

\bibitem{virgo:2015}
F~Acernese et~al.
\newblock Advanced {V}irgo: a second-generation interferometric gravitational wave detector.
\newblock {\em Classical and Quantum Gravity}, 32(2):024001, dec 2014.

\bibitem{kagra:2020}
T~Akutsu et~al.
\newblock {Overview of {KAGRA}: {D}etector design and construction history}.
\newblock {\em Progress of Theoretical and Experimental Physics}, 2021(5):05A101, 08 2020.

\bibitem{ligo_ccsn:2020}
B.~P. Abbott et~al.
\newblock Optically targeted search for gravitational waves emitted by core-collapse supernovae during the first and second observing runs of advanced {LIGO} and advanced {V}irgo.
\newblock {\em Phys. Rev. D}, 101:084002, Apr 2020.

\bibitem{vartanyan:2023b}
David Vartanyan, Adam Burrows, Tianshu Wang, Matthew S.~B. Coleman, and Christopher~J. White.
\newblock The gravitational-wave signature of core-collapse supernovae.
\newblock {\em arXiv preprint arXiv:2302.07092}, 2023.

\bibitem{schnetter:2007}
Erik Schnetter, Christian~D. Ott, Gabrielle Allen, Peter Diener, Tom Goodale, Thomas Radke, Edward Seidel, and John Shalf.
\newblock Cactus framework: Black holes to gamma ray bursts.
\newblock {\em arXiv preprint arXiv:0707.1607}, 2007.

\bibitem{ott:2009}
Christian~D Ott.
\newblock The gravitational-wave signature of core-collapse supernovae.
\newblock {\em Classical and Quantum Gravity}, 26(6):063001, feb 2009.

\bibitem{klimenko:2016}
S.~Klimenko, G.~Vedovato, M.~Drago, F.~Salemi, V.~Tiwari, G.~A. Prodi, C.~Lazzaro, K.~Ackley, S.~Tiwari, C.~F. Da~Silva, and G.~Mitselmakher.
\newblock Method for detection and reconstruction of gravitational wave transients with networks of advanced detectors.
\newblock {\em Phys. Rev. D}, 93:042004, Feb 2016.

\bibitem{cornish:2015}
Neil~J Cornish and Tyson~B Littenberg.
\newblock Bayeswave: {B}ayesian inference for gravitational wave bursts and instrument glitches.
\newblock {\em Classical and Quantum Gravity}, 32(13):135012, jun 2015.

\bibitem{meyer:2022}
Renate Meyer, Matthew~C. Edwards, Patricio Maturana-Russel, and Nelson Christensen.
\newblock Computational techniques for parameter estimation of gravitational wave signals.
\newblock {\em WIREs Computational Statistics}, 14(1):e1532, 2022.

\bibitem{heng:2009}
Ik~Siong Heng.
\newblock Rotating stellar core-collapse waveform decomposition: a principal component analysis approach.
\newblock {\em Classical and Quantum Gravity}, 26(10):105005, apr 2009.

\bibitem{roever:2009}
Christian R\"over, Marie-Anne Bizouard, Nelson Christensen, Harald Dimmelmeier, Ik~Siong Heng, and Renate Meyer.
\newblock Bayesian reconstruction of gravitational wave burst signals from simulations of rotating stellar core collapse and bounce.
\newblock {\em Phys. Rev. D}, 80:102004, Nov 2009.

\bibitem{logue:2012}
J.~Logue, C.~D. Ott, I.~S. Heng, P.~Kalmus, and J.~H.~C. Scargill.
\newblock Inferring core-collapse supernova physics with gravitational waves.
\newblock {\em Phys. Rev. D}, 86:044023, Aug 2012.

\bibitem{edwards:2014}
Matthew~C. Edwards, Renate Meyer, and Nelson Christensen.
\newblock Bayesian parameter estimation of core collapse supernovae using gravitational wave simulations.
\newblock {\em Inverse Problems}, 30(11):114008, oct 2014.

\bibitem{bizouard:2021}
Marie-Anne Bizouard, Patricio Maturana-Russel, Alejandro Torres-Forn\'e, Martin Obergaulinger, Pablo Cerd\'a-Dur\'an, Nelson Christensen, Jos\'e~A. Font, and Renate Meyer.
\newblock Inference of protoneutron star properties from gravitational-wave data in core-collapse supernovae.
\newblock {\em Phys. Rev. D}, 103:063006, Mar 2021.

\bibitem{bizouard:2023}
Tristan Bruel, Marie-Anne Bizouard, Martin Obergaulinger, Patricio Maturana-Russel, Alejandro Torres-Forn\'e, Pablo Cerd\'a-Dur\'an, Nelson Christensen, Jos\'e~A. Font, and Renate Meyer.
\newblock Inference of protoneutron star properties in core-collapse supernovae from a gravitational-wave detector network.
\newblock {\em Phys. Rev. D}, 107:083029, Apr 2023.

\bibitem{astone:2018}
P.~Astone, P.~Cerd\'a-Dur\'an, I.~Di~Palma, M.~Drago, F.~Muciaccia, C.~Palomba, and F.~Ricci.
\newblock New method to observe gravitational waves emitted by core collapse supernovae.
\newblock {\em Phys. Rev. D}, 98:122002, Dec 2018.

\bibitem{lopez:2021}
M.~L\'opez, I.~Di~Palma, M.~Drago, P.~Cerd\'a-Dur\'an, and F.~Ricci.
\newblock Deep learning for core-collapse supernova detection.
\newblock {\em Phys. Rev. D}, 103:063011, Mar 2021.

\bibitem{iess:2023}
{Iess, Alberto}, {Cuoco, Elena}, {Morawski, Filip}, {Nicolaou, Constantina}, and {Lahav, Ofer}.
\newblock {LSTM} and {CNN} application for core-collapse supernova search in gravitational wave real data.
\newblock {\em Astronomy and Astrophysics}, 669:A42, 2023.

\bibitem{chan:2020}
Man~Leong Chan, Ik~Siong Heng, and Chris Messenger.
\newblock Detection and classification of supernova gravitational wave signals: A deep learning approach.
\newblock {\em Phys. Rev. D}, 102:043022, Aug 2020.

\bibitem{iess:2020}
Alberto Iess, Elena Cuoco, Filip Morawski, and Jade Powell.
\newblock Core-collapse supernova gravitational-wave search and deep learning classification.
\newblock {\em Machine Learning: Science and Technology}, 1(2):025014, may 2020.

\bibitem{edwards:2021}
Matthew~C. Edwards.
\newblock Classifying the equation of state from rotating core collapse gravitational waves with deep learning.
\newblock {\em Phys. Rev. D}, 103:024025, Jan 2021.

\bibitem{goodfellow:2016}
Ian~J. Goodfellow, Yoshua Bengio, and Aaron Courville.
\newblock {\em Deep Learning}.
\newblock MIT Press, Cambridge, MA, USA, 2016.
\newblock \url{http://www.deeplearningbook.org}.

\bibitem{kingma:2022}
Diederik~P Kingma and Max Welling.
\newblock Auto-encoding variational {B}ayes.
\newblock {\em arXiv preprint arXiv:1312.6114}, 2022.

\bibitem{vaswani:2017}
Ashish Vaswani, Noam Shazeer, Niki Parmar, Jakob Uszkoreit, Llion Jones, Aidan~N. Gomez, Lukasz Kaiser, and Illia Polosukhin.
\newblock Attention is all you need.
\newblock {\em arXiv preprint arXiv:1706.03762}, 2017.

\bibitem{radford:2018}
A.~Radford, K.~Narasimhan, T.~Salimans, and I.~Sutskever.
\newblock Improving language understanding by generative pre-training, 2018.
\newblock Accessed: 2024-02-09.

\bibitem{shi:2020}
Zhan Shi, Xu~Zhou, Xipeng Qiu, and Xiaodan Zhu.
\newblock Improving image captioning with better use of captions.
\newblock {\em arXiv preprint arXiv:2006.11807}, 2020.

\bibitem{rombach:2022}
Robin Rombach, Andreas Blattmann, Dominik Lorenz, Patrick Esser, and Björn Ommer.
\newblock High-resolution image synthesis with latent diffusion models.
\newblock In {\em 2022 IEEE/CVF Conference on Computer Vision and Pattern Recognition (CVPR)}, pages 10674--10685, 2022.

\bibitem{jorgensen:2018}
Peter~B. Jørgensen, Mikkel~N. Schmidt, and Ole Winther.
\newblock Deep generative models for molecular science.
\newblock {\em Molecular Informatics}, 37(1-2):1700133, 2018.

\bibitem{ingraham:2019}
John Ingraham, Vikas Garg, Regina Barzilay, and Tommi Jaakkola.
\newblock Generative models for graph-based protein design.
\newblock In H.~Wallach, H.~Larochelle, A.~Beygelzimer, F.~d\textquotesingle Alch\'{e}-Buc, E.~Fox, and R.~Garnett, editors, {\em Advances in Neural Information Processing Systems}, volume~32. Curran Associates, Inc., 2019.

\bibitem{anand:2018}
Namrata Anand and Possu Huang.
\newblock Generative modeling for protein structures.
\newblock In S.~Bengio, H.~Wallach, H.~Larochelle, K.~Grauman, N.~Cesa-Bianchi, and R.~Garnett, editors, {\em Advances in Neural Information Processing Systems}, volume~31. Curran Associates, Inc., 2018.

\bibitem{ramezanian:2022}
Mahta Ramezanian-Panahi, Germán Abrevaya, Jean-Christophe Gagnon-Audet, Vikram Voleti, Irina Rish, and Guillaume Dumas.
\newblock Generative models of brain dynamics.
\newblock {\em Frontiers in Artificial Intelligence}, 5, 2022.

\bibitem{schawinski:2017}
Kevin Schawinski, Ce~Zhang, Hantian Zhang, Lucas Fowler, and Gokula~Krishnan Santhanam.
\newblock {Generative adversarial networks recover features in astrophysical images of galaxies beyond the deconvolution limit}.
\newblock {\em Monthly Notices of the Royal Astronomical Society: Letters}, 467(1):L110--L114, 01 2017.

\bibitem{benedetto:2023}
Vincenzo Benedetto, Francesco Gissi, Gioele Ciaparrone, and Luigi Troiano.
\newblock {AI} in gravitational wave analysis, an overview.
\newblock {\em Applied Sciences}, 13(17), 2023.

\bibitem{green:2020}
Stephen~R. Green, Christine Simpson, and Jonathan Gair.
\newblock Gravitational-wave parameter estimation with autoregressive neural network flows.
\newblock {\em Phys. Rev. D}, 102:104057, Nov 2020.

\bibitem{gabbard:2022}
H.~Gabbard, C.~Messenger, I.S. Heng, F.~Tonolini, and R.~Murray-Smith.
\newblock Bayesian parameter estimation using conditional variational autoencoders for gravitational-wave astronomy.
\newblock {\em Nature Physics}, 18:112--117, 2022.

\bibitem{liao:2021}
Chung-Hao Liao and Feng-Li Lin.
\newblock Deep generative models of gravitational waveforms via conditional autoencoder.
\newblock {\em Phys. Rev. D}, 103:124051, Jun 2021.

\bibitem{lee:2021}
Joongoo Lee, Sang~Hoon Oh, Kyungmin Kim, Gihyuk Cho, John~J. Oh, Edwin~J. Son, and Hyung~Mok Lee.
\newblock Deep learning model on gravitational waveforms in merging and ringdown phases of binary black hole coalescences.
\newblock {\em Phys. Rev. D}, 103:123023, Jun 2021.

\bibitem{mcginn:2021}
J~McGinn, C~Messenger, M~J Williams, and I~S Heng.
\newblock Generalised gravitational wave burst generation with generative adversarial networks.
\newblock {\em Classical and Quantum Gravity}, 38(15):155005, jun 2021.

\bibitem{lopez:2022}
Melissa Lopez, Vincent Boudart, Kerwin Buijsman, Amit Reza, and Sarah Caudill.
\newblock Simulating transient noise bursts in {LIGO} with generative adversarial networks.
\newblock {\em Phys. Rev. D}, 106:023027, Jul 2022.

\bibitem{powell:2023}
Jade Powell, Ling Sun, Katinka Gereb, Paul~D Lasky, and Markus Dollmann.
\newblock Generating transient noise artefacts in gravitational-wave detector data with generative adversarial networks.
\newblock {\em Classical and Quantum Gravity}, 40(3):035006, jan 2023.

\bibitem{chua:2019}
Alvin J.~K. Chua, Chad~R. Galley, and Michele Vallisneri.
\newblock Reduced-order modeling with artificial neurons for gravitational-wave inference.
\newblock {\em Phys. Rev. Lett.}, 122:211101, May 2019.

\bibitem{chua:2021}
Alvin J.~K. Chua, Michael~L. Katz, Niels Warburton, and Scott~A. Hughes.
\newblock Rapid generation of fully relativistic extreme-mass-ratio-inspiral waveform templates for {LISA} data analysis.
\newblock {\em Phys. Rev. Lett.}, 126:051102, Feb 2021.

\bibitem{katz:2021}
Michael~L. Katz, Alvin J.~K. Chua, Lorenzo Speri, Niels Warburton, and Scott~A. Hughes.
\newblock Fast extreme-mass-ratio-inspiral waveforms: {N}ew tools for millihertz gravitational-wave data analysis.
\newblock {\em Phys. Rev. D}, 104:064047, Sep 2021.

\bibitem{grimbergen:2024}
Tim Grimbergen, Stefano Schmidt, Chinmay Kalaghatgi, and Chris van~den Broeck.
\newblock Generating higher order modes from binary black hole mergers with machine learning.
\newblock {\em arXiv preprint arXiv:2402.06587}, 2024.

\bibitem{basodi:2020}
Sunitha Basodi, Chunyan Ji, Haiping Zhang, and Yi~Pan.
\newblock Gradient amplification: An efficient way to train deep neural networks.
\newblock {\em arXiv preprint arXiv:2006.10560}, 2020.

\bibitem{arjovsky:2017}
Martin Arjovsky, Soumith Chintala, and Léon Bottou.
\newblock Wasserstein {GAN}.
\newblock {\em arXiv preprint arXiv:1701.07875}, 2017.

\bibitem{metz:2016}
Luke Metz, Ben Poole, David Pfau, and Jascha Sohl-Dickstein.
\newblock Unrolled generative adversarial networks.
\newblock {\em arXiv preprint arXiv:1611.02163}, 2016.

\bibitem{salimans:2016}
Tim Salimans, Ian~J. Goodfellow, Wojciech Zaremba, Vicki Cheung, Alec Radford, and Xi~Chen.
\newblock Improved techniques for training {GAN}s.
\newblock {\em arXiv preprint arXiv:1606.03498}, 2016.

\bibitem{szegedy:2015}
Christian Szegedy, Vincent Vanhoucke, Sergey Ioffe, Jonathon Shlens, and Zbigniew Wojna.
\newblock Rethinking the inception architecture for computer vision.
\newblock {\em arXiv preprint arXiv:1512.00567}, 2015.

\bibitem{ioeff:2015}
Sergey Ioffe and Christian Szegedy.
\newblock Batch normalization: Accelerating deep network training by reducing internal covariate shift.
\newblock {\em arXiv preprint arXiv:1502.03167}, 2015.

\bibitem{richers-data:2016}
S.~Richers, C.~D. Ott, E.~Abdikamalov, E.~O'Connor, and C.~Sullivan.
\newblock Equation of state effects on gravitational waves from rotating core collapse [data set], 2016.

\bibitem{mitra:2024}
A~Mitra, D~Orel, Y~S Abylkairov, B~Shukirgaliyev, and E~Abdikamalov.
\newblock {Probing nuclear physics with supernova gravitational waves and machine learning}.
\newblock {\em Monthly Notices of the Royal Astronomical Society}, page stae714, 03 2024.

\bibitem{smith:1984}
J.~Smith and P.~Gossett.
\newblock A flexible sampling-rate conversion method.
\newblock In {\em ICASSP '84. IEEE International Conference on Acoustics, Speech, and Signal Processing}, volume~9, pages 112--115, 1984.

\bibitem{dcgans_pytorch_tutorial}
{Pytorch team}.
\newblock Dcgans tutorial.
\newblock \url{https://pytorch.org/tutorials/beginner/dcgan_faces_tutorial.html}.
\newblock Accessed: 2023-07-15.

\bibitem{borji:2022}
Ali Borji.
\newblock Pros and cons of {GAN} evaluation measures: New developments.
\newblock {\em Computer Vision and Image Understanding}, 215:103329, 2022.

\bibitem{heusel:2017}
Martin Heusel, Hubert Ramsauer, Thomas Unterthiner, Bernhard Nessler, G{\"{u}}nter Klambauer, and Sepp Hochreiter.
\newblock {GAN}s trained by a two time-scale update rule converge to a nash equilibrium.
\newblock {\em CoRR}, abs/1706.08500, 2017.

\bibitem{mirza:2014}
Mehdi Mirza and Simon Osindero.
\newblock Conditional generative adversarial nets.
\newblock {\em CoRR}, abs/1411.1784, 2014.

\bibitem{esteban:2017}
Cristóbal Esteban, Stephanie~L. Hyland, and Gunnar Rätsch.
\newblock Real-valued (medical) time series generation with recurrent conditional {GAN}s.
\newblock {\em arXiv preprint arXiv:1706.02633}, 2017.

\bibitem{cho:2014}
Kyunghyun Cho, Bart van Merrienboer, Dzmitry Bahdanau, and Yoshua Bengio.
\newblock On the properties of neural machine translation: Encoder-decoder approaches.
\newblock {\em arXiv preprint arXiv:1409.1259}, 2014.

\bibitem{chung:2014}
Junyoung Chung, Caglar Gulcehre, KyungHyun Cho, and Yoshua Bengio.
\newblock Empirical evaluation of gated recurrent neural networks on sequence modeling.
\newblock {\em arXiv preprint arXiv:1412.3555}, 2014.

\bibitem{yang:2020}
Shudong Yang, Xueying Yu, and Ying Zhou.
\newblock Lstm and gru neural network performance comparison study: Taking yelp review dataset as an example.
\newblock In {\em 2020 International Workshop on Electronic Communication and Artificial Intelligence (IWECAI)}, pages 98--101, 2020.

\bibitem{wen:2023}
Qingsong Wen, Tian Zhou, Chaoli Zhang, Weiqi Chen, Ziqing Ma, Junchi Yan, and Liang Sun.
\newblock Transformers in time series: A survey.
\newblock {\em arXiv preprint arXiv:2202.07125}, 2023.

\bibitem{fu:2023a}
Daniel~Y. Fu, Elliot~L. Epstein, Eric Nguyen, Armin~W. Thomas, Michael Zhang, Tri Dao, Atri Rudra, and Christopher Ré.
\newblock Simple hardware-efficient long convolutions for sequence modeling.
\newblock {\em arXiv preprint arXiv:2302.06646}, 2023.

\bibitem{fu:2023b}
Daniel~Y. Fu, Tri Dao, Khaled~K. Saab, Armin~W. Thomas, Atri Rudra, and Christopher Ré.
\newblock Hungry hungry hippos: Towards language modeling with state space models.
\newblock {\em arXiv preprint arXiv:2212.14052}, 2023.

\bibitem{poli:2023}
Michael Poli, Stefano Massaroli, Eric Nguyen, Daniel~Y. Fu, Tri Dao, Stephen Baccus, Yoshua Bengio, Stefano Ermon, and Christopher Ré.
\newblock Hyena hierarchy: Towards larger convolutional language models.
\newblock {\em arXiv preprint arXiv:2302.10866}, 2023.

\bibitem{xu:2023}
Lingling Xu, Haoran Xie, Si-Zhao~Joe Qin, Xiaohui Tao, and Fu~Lee Wang.
\newblock Parameter-efficient fine-tuning methods for pretrained language models: A critical review and assessment.
\newblock {\em arXiv preprint arXiv:2312.12148}, 2023.

\end{thebibliography}

\end{document}